\documentclass[english,prb,showpacs,amsmath,amssymb,aps,12pt,superscriptaddress]{revtex4}
\pdfoutput=1

\usepackage{amsfonts}
\usepackage{amssymb}
\usepackage{bm}
\input epsf

\usepackage[latin9]{inputenc}
\usepackage[english]{babel}
\usepackage{esint}
\usepackage{braket}

\usepackage[caption=false]{subfig}
\usepackage[absolute,overlay]{textpos}
\usepackage{graphicx}
\usepackage{epstopdf} 
\usepackage{pdfsync}
\usepackage[dvipsnames]{xcolor}
\usepackage[normalem]{ulem}
\usepackage{hyperref}
\usepackage{slashed}
\usepackage{dsfont}
\usepackage{float}

\newcommand{\be}[1]{ \begin{eqnarray} \mbox{$\label{#1}$} }
   
\newcommand{\ee}{\end{eqnarray}}
\newcommand{\eeq}{\end{equation}}

\newcommand\oncite [1] {Ref.\,\onlinecite{#1} }

\begin{document}
\title{A field theory approach to Breit-type Hamiltonians \\ in gapped Dirac systems}

\author{Xinhong Zhou}
\affiliation{Tsung-Dao Lee Institute, Shanghai Jiao Tong University, Shanghai, 201210, China}
\author{T. H. Hansson}
\affiliation{Department of Physics, Stockholm University, AlbaNova University Center, 106 91 Stockholm, Sweden}
\author{Varsha Subramanyan}
\affiliation{Tsung-Dao Lee Institute, Shanghai Jiao Tong University, Shanghai, 201210, China}
\author{Qing-Dong Jiang}
\affiliation{Tsung-Dao Lee Institute, Shanghai Jiao Tong University, Shanghai, 201210, China}
\begin{abstract}
We develop a path-integral-based field theory for deriving Breit-type low-energy Hamiltonians for gapped Dirac systems coupled to both vector and axial gauge fields. Treating the mass gap as the large energy scale, we integrate out the high-energy component of the Dirac spinor and  obtain a canonical Schr\"odinger description for the remaining low-energy degrees of freedom. For the ordinary Dirac equation, our method reproduces the conventional Breit Hamiltonian order by order. In the presence of axial gauge fields, however, the resulting Hamiltonian contains additional vector--axial couplings that have no analogue in the traditional electromagnetic case. We illustrate the method using three- and two-dimensional Dirac models and discuss its relevance to gapped Weyl systems, where dynamical axial fields can generate distinctive low-energy transport signatures.
\end{abstract}
                                        
\maketitle   

\section{Introduction}

An early triumph of Dirac's theory of the electron was the emergence of its low-energy, or equivalently large-mass, limit. In this limit the relativistic Dirac equation reduces to an effective non-relativistic Hamiltonian containing the familiar Schr\"odinger kinetic energy, the Zeeman coupling, and relativistic corrections such as the Darwin and spin-orbit terms \cite{Nolting2017}. For two-electron systems, related corrections were first derived by Breit in 1929, and were later placed in a systematic framework by Foldy and Wouthuysen, who used a sequence of unitary transformations to decouple the positive- and negative-energy components of the Dirac spinor~\cite{Silenko2008}. The resulting Breit-type Hamiltonian is organized as an expansion in inverse powers of the mass and provides one of the earliest examples of a controlled low-energy effective Hamiltonian derived from a relativistic theory.

Dirac-type Hamiltonians are now ubiquitous in condensed matter physics. They appear, for example, in the effective descriptions of topological insulators \cite{TIText}, Weyl and Dirac semimetals \cite{Armitage2018}, graphene-like systems \cite{MIRANSKY20151}, and more general gapped band structures. In these systems, the ``mass'' need not be the bare electron mass. Instead, it often represents a band gap generated by spin-orbit coupling, sublattice symmetry breaking, excitonic order, or other microscopic mechanisms \cite{Zhangshoucheng2013,Roy2015,Yu2021}. This observation suggests that the logic behind the Breit Hamiltonian should have a broader condensed-matter counterpart: given a gapped Dirac-like system, one should be able to systematically integrate out the high-energy bands and obtain a low-energy Hamiltonian for the remaining degrees of freedom.

In this work we develop such a method for Dirac models coupled not only to ordinary vector gauge fields, but also to axial gauge fields. The latter arise naturally in Weyl and Dirac materials, where strain or order-parameter textures can couple with opposite signs to fermions of opposite chirality\cite{Cortijo2015}. A representative model of interest is a massive Dirac theory with vector potential $A_\mu$, axial potential $\tilde A_\mu^5$, Weyl-node separation $b_\mu$, and a complex mass $m e^{i\theta(x)}$. Here $m$ sets the size of the gap, while the phase $\theta(x)$ may describe, for instance, the phase of an excitonic order parameter in a gapped Weyl system~\cite{Zyuzin2012,Yu2021}. Such Dirac models also play an important role in the classification and response theory of gapped topological phases~\cite{Schnyder2008,Ryu2010}.

Our goal is to derive the corresponding Breit-type Hamiltonian by using path integral methods. The strategy is conceptually simple. We separate the Dirac spinor into low- and high-energy components, integrate out the high-energy component, and then expand the resulting theory in inverse powers of the gap. For the ordinary Dirac equation, this procedure reproduces the conventional Breit Hamiltonian order by order. For Dirac systems coupled to axial gauge fields, however, the same procedure generates additional terms that have no analogue in the standard electromagnetic problem. These terms encode the combined effects of the mass gap, vector gauge fields, and axial gauge fields, and are therefore relevant for the low-energy dynamics of gapped Weyl and Dirac materials.

The field theory formulation has two advantages. First, it makes clear that the Breit Hamiltonian is naturally understood as a low-energy effective Hamiltonian obtained by integrating out negative-energy degrees of freedom. This viewpoint parallels the logic of effective field theory and the renormalization group, which were not available in the original derivations of Breit and Foldy--Wouthuysen. Second, the method is readily generalized to multi-band systems and to external fields with spatial and temporal dependence. Although in this paper we write explicit expressions only to low orders in the inverse gap, the procedure can be systematically extended to higher orders in the external fields and their derivatives.

We emphasize that our aim is to construct effective low-energy \emph{Hamiltonians}, rather than only effective actions. In many field theory approaches, integrating out high-energy modes naturally produces an effective action, often containing higher-order time derivatives. Here we use the part of the effective action that is first order in time derivatives to identify the appropriate symplectic form. This allows us to rescale the low-energy fields so that they obey canonical equal-time anticommutation relations, and hence yields a genuine effective Hamiltonian. This step is essential for connecting the path integral derivation to the canonical Hamiltonian framework commonly used in condensed matter physics.

The paper is structured as follows. In Sec.~\ref{S2}, we first remove the phase of the complex mass by a chiral rotation, which produces the usual anomaly-related topological term and gives a massive Dirac theory coupled to an effective axial potential. In Sec.~\ref{General process of path integral method}, we present the general path integral method: we integrate out the high-energy component, identify the modified symplectic structure, rescale the low-energy field, and derive the generalized Breit-type Hamiltonian. This method is then illustrated with two examples, namely three- and two-dimensional gapped Dirac systems. In Sec.~\ref{discussionandsummary}, we discuss potential transport signatures arising from dynamical axial fields. In particular, we identify strained Weyl semimetals with an exciton-induced gap as promising candidates where our effective Hamiltonian may yield observable effects, as further explored in our companion work~\cite{shortpaper}.

\section{Chiral rotation of the $\theta(x)$ term}\label{S2}

We begin with the $(3+1)d$ Dirac Lagrangian
\be{dirac}\label{L3D}
{\cal L}=\bar\psi\left\{i\gamma^\mu[\partial_\mu+iA_\mu+i\gamma^5({\tilde{A}^5_\mu}+b_\mu)]-me^{i\gamma^5\theta(x)}\right\}\psi,
\ee
where $A_\mu$ denotes the external vector potential, $\tilde{A}^5_\mu$ are external axial potential which can be generated by, for instance, acoustic  means~\cite{Ilan2020}, and $b_\mu$ characterizes the separation of the Weyl nodes in energy-momentum space. The complex mass parameter $m e^{i\theta(x)}$ opens a gap in the Dirac spectrum. In condensed matter systems, such a term may describe, for example, a Weyl semimetal in which the left- and right-handed Weyl nodes are coupled by an excitonic order parameter~\cite{Grigoreva2025,Chang2021}. 
To make the excitonic interpretation explicit, the complex mass term can be rewritten as~\cite{Zyuzin2012,Yu2021}
\be{How to see the exciton pairing in the Lagrangian}
\bar\psi[-me^{i\gamma^5\theta(x)}]\psi=-[me^{i\theta(x)}\psi_L^\dagger\psi_R+h.c.]
\ee

Thus, the mass term mixes the left- and right-handed fermions, which are defined by the chiral projection operators $P_{\pm}=(1\pm\gamma^5)/2$. Before integrating out the negative-energy degrees of freedom, it is useful to remove the phase of the complex mass by performing the chiral rotation
\be{chiral rotation to simplify the origin Lagrangian}
\psi\rightarrow\psi'&=& e^{-i\gamma^5\frac{\theta(x)}{2}}\psi\\
\bar\psi\rightarrow\bar\psi'&=&\bar\psi e^{-i\gamma^5\frac{\theta(x)}{2}}
\ee
At the classical level, this transformation makes the mass real and transfers the phase gradient into the axial gauge sector. At the quantum level, however, the chiral rotation also changes the path-integral measure. The resulting nontrivial Jacobian ${\cal J}_{chiral}$ is due to the axial anomaly and gives the familiar $\theta$ term~\cite{Zyuzin2012,Huang2017,Grigoreva2025}. It can be proven that  ${\cal J}_{chiral}$ takes the form
\be{Jacobian for the chiral rotation in this problem}
{\cal J}_{chiral}&=&\mathrm{exp}\left\{i\int_x \theta(x)\left[-\frac{\epsilon^{\mu\nu\alpha\beta}}{64\pi^2}(F^+_{\mu\nu}F^+_{\alpha\beta}+F^-_{\mu\nu}F^-_{\alpha\beta})\right]\right\}
\ee
which is independent of the mass $m$. The chirally rotated action can therefore be written as~\cite{Huang2017,Yu2021}
\be{equivalence action to describe the system}
\Gamma[A_\mu,{A^5_\mu}]&=&\int_x\left\{-\frac{\theta(x)}{64\pi^2}\epsilon^{\mu\nu\alpha\beta}(F_{\mu\nu}^{+}F_{\alpha\beta}^{+}+F_{\mu\nu}^{-}F_{\alpha\beta}^{-})\right\}\nonumber\\
&+&\int_x\bar\psi[i\gamma^\mu(\partial_\mu+iA_\mu+i\gamma^5 {A^5_\mu})-m]\psi\label{postrot}
\ee
where the effective axial potential ${A^5_\mu}(x)$ and the field strengths $F_{\mu\nu}^\pm$ are defined by
\be{definition of S_mu and F_munu}
{A^5_\mu}(x)&=&\tilde {A^5_\mu}(x)+b_\mu-\frac{\partial_\mu\theta(x)}{2}\label{modification of S_mu from exciton coupling}\\
F_{\mu\nu}^\pm&=&\partial_\mu(A_\nu\pm {A^5_\nu})-\partial_\nu(A_\mu\pm {A^5_\mu}).
\ee
The first term in Eq.~\eqref{equivalence action to describe the system} is topological in origin and a total derivative. It is important in understanding the bulk-boundary correspondence and the topological response of Weyl semimetals, and is crucial in explaining axion electrodynamics, dynamical piezomagnetic effects and so on~\cite{Rylands2021,Grigoreva2025,Yu2021,Liebman2025}. In the present work, however, we focus on the bulk low-energy Hamiltonian obtained from the mass-dependent part of the theory. We therefore concentrate on the second term in Eq.~\eqref{equivalence action to describe the system}, which is a massive Dirac Lagrangian coupled to both the vector potential $A_\mu$ and the effective axial potential ${A^5_\mu}$. This term provides the starting point for the large-mass expansion developed below.

\section{The path integral technique}\label{General process of path integral method}

We now present the general path-integral procedure for deriving the low-energy Hamiltonian. We consider a {\color{black}Dirac-type} fermionic field $\psi=(\phi,\chi)^{\mathrm{T}}$, where $\phi$ and $\chi$ denote the positive- and negative-energy components, respectively. The two sectors are separated by a large gap $2m$. In this basis, the Lagrangian can be recast as
\be{general 2n band Lagrangian we calculate}
{\cal L}&=&\psi^\dagger\begin{pmatrix}
    i{\cal D}_0& {\cal D}^\dagger\\ {\cal D}&i{\cal D}_0+2m
\end{pmatrix}\psi=\phi^\dagger(i{\cal D}_0)\phi+\chi^\dagger(i{\cal D}_0+2m)\chi+\phi^\dagger{\cal D}^\dagger\chi+\chi^\dagger{\cal D}\phi\label{L2n}.
\ee
Here ${\cal D}_0=\partial_0+iA_0+iV$, where $V$ is an effective potential independent of $A_0$, and ${\cal D}$ contains the momentum-dependent couplings between the positive- and negative-energy sectors. This form applies quite generally to a $2N$-band {\color{black}Dirac-type} fermionic system {in (d+1) dimensional spacetime} with a large gap $2m$ separating the two sets of bands. While the explicit form of ${\cal D}$ may depend on the dimensionality and other specifics of the system, the path-integral method can be carried out in this general notation.

To integrate out $\chi$ and organize the result as an expansion in inverse powers of $m$, we assume without loss of generality that the Fermi surface lies close to the positive-energy band edge, so that the relevant momenta satisfy $|\mathbf{p}|\ll m$. After integrating out the negative energy part $\chi$ and perform order expansion respect to band gap $2m$, we reach the following  effective Lagrangian\footnote{See similar idea in high density effective field theory\cite{HONG2000118,SCHAFER2003251,Son2013} where one is interested in the particles near the Fermi surface so order counting respect to chemical potential $O(\mu^{-n}),n=1,2...$ is performed to get the effective Lagrangian instead of Hamiltonian.}:
\be{n band effective Lagrangian before rescaling}
{\cal L}_{eff}&=&\phi^\dagger[i{\cal D}_0-{\cal D}^\dagger(i{\cal D}_0+2m)^{-1}{\cal D}]\phi\nonumber\\
&\simeq&\phi^\dagger\left[i{\cal D}_0-\frac{1}{2m}{\cal D}^\dagger{\cal D}+\frac{1}{4m^2}{\cal D}^\dagger(i{\cal D}_0){\cal D}\right]\phi\label{Leff_gen}
\ee
From the Lagrangian, the density operator is given by:
\be{calculation the density operator}
\rho=-\frac{\delta{\cal L}_{eff}}{\delta A_0}=\phi^\dagger\left(1+\frac{1}{4m^2}{\cal D}^\dagger{\cal D}\right)\phi.
\ee
Clearly this is different from the Schr\"odinger expression $\rho = \tilde\phi^\dagger\tilde\phi$, so we cannot directly interpret $\phi$ as a Schr\"odinger field. Therefore, it is useful to rescale the fields appropriately to identify the Schrodinger equation, and thus the effective Hamiltonian. To this end, we draw attention to the symplectic term in the Lagrangian: 
\be{sympletic form in the effective Lagrangian}
{\cal L}_{sym}=\phi^\dagger\left[i\partial_0+\frac{1}{4m^2}{\cal D}^\dagger(i\partial_0){\cal D}\right]\phi
\ee 
This term encodes the canonical commutation relations of the relevant fields and allows us to formulate a conventional Schr\"odinger dynamics. We have only used the terms with one time derivative, i.e. $\sim \partial_t$, to define the Lagrangian, and hence, the symplectic form. Terms with $\partial_t^2$ and higher should be treated perturbatively\cite{Donoghue}, but see also e.g. \oncite{Knetter1994}. For the fermionic fields here, we can read off the equal time anticommutator
\be{equal time anticommutator for L_(sym)}
[\phi(t,\mathbf{x}),\phi^\dagger(t,\mathbf{y})]_{+}=\left[1+\frac{1}{4m^2}{\cal D}^\dagger{\cal D}\right]^{-1}\delta^d(\mathbf{x}-\mathbf{y})
\ee
However, one could rescale the fields as
\be{rescale the positive energy fermionic field}
\tilde\phi^\dagger&=&\phi^\dagger\left(1+\frac{1}{4m^2}{\cal D}^\dagger{\cal D}\right)^{\frac{1}{2}}\label{general rescale field 1}\\
\tilde\phi&=&\left(1+\frac{1}{4m^2}{\cal D}^\dagger{\cal D}\right)^{\frac{1}{2}}\phi\label{general rescale field 2}
\ee
and obtain the canonical anticommutation relationship
\be{canonical anticommutation relationship after field rescaling}
[\tilde\phi(t,\mathbf{x}),\tilde\phi^\dagger(t,\mathbf{y})]_{+}=\delta^d(\mathbf{x}-\mathbf{y}).
\ee
Now we can write the effective Lagrangian ${\cal L}_{eff}$ respect to rescaled fields $(\tilde\phi^\dagger,\tilde\phi)$ to obtain the Schr\"odinger form $(i\partial_t-\hat H)\tilde\phi=0$ of the equations of motion up to $O(m^{-2})$:
\be{n band effective Lagrangian after rescaling}
{\cal L}_{eff}&=&\tilde\phi^\dagger\left(1-\frac{1}{8m^2}{\cal D}^\dagger{\cal D}\right)\left[i{\cal D}_0-\frac{1}{2m}{\cal D}^\dagger{\cal D}+\frac{1}{4m^2}{\cal D}^\dagger(i{\cal D}_0){\cal D}\right]\left(1-\frac{1}{8m^2}{\cal D}^\dagger{\cal D}\right)\tilde\phi\nonumber\\
&\simeq&\tilde\phi^\dagger\left\{i{\cal D}_0-\frac{1}{2m}{\cal D}^\dagger{\cal D}+\frac{1}{8m^2}[2{\cal D}^\dagger(i{\cal D}_0){\cal D}-{\cal D}^\dagger{\cal D}(i{\cal D}_0)-(i{\cal D}_0){\cal D}^\dagger{\cal D}]\right\}\tilde\phi\nonumber\\
&=&\tilde\phi^\dagger\left\{i{\cal D}_0-\frac{1}{2m}{\cal D}^\dagger{\cal D}+\frac{1}{8m^2}({\cal D}^\dagger[i{\cal D}_0,{\cal D}]_{-}+h.c.)\right\}\tilde\phi
\ee
The effective $N$-bands Hamiltonian is given by $\hat H_{eff}=\sum_n\hat H_{eff}^{(n)}$ where $n=0,1,2$ denotes the effective Hamiltonian of $O(m^{-n})$ order:
\be{final expression of general effective Hamiltonian}
\hat H_{eff}=(A_0+V)+\frac{1}{2m}({\cal D}^\dagger{\cal D})-\frac{1}{8m^2}({\cal D}^\dagger[i{\cal D}_0,{\cal D}]_{-}+h.c.).
\ee
We have thus obtained the generalized Breit Hamiltonian for a {\color{black}Dirac-type} $2N$-band system. It is worth emphasizing that the field rescaling dictated by the symplectic structure is essential: without this step, the path-integral effective action does not directly lead to a canonical Schr\"odinger Hamiltonian for the low-energy degrees of freedom. {\color{black} We note that Eq.~(18) is equivalent to the Foldy--Wouthuysen Hamiltonian (up to the order $O(m^{-2})$ as shown below). However, the path-integral formulation provides a more systematic and readily generalizable route: by directly integrating out the negative-energy modes, it avoids the order-by-order construction of model-dependent unitary transformations and naturally accommodates additional ingredients, such as axial gauge potentials, within the same framework.}

\subsection{The Breit Hamiltonian for $(3+1)d$ gapped Dirac systems}
The general result of Eq.\eqref{final expression of general effective Hamiltonian} can now be applied to the mass-dependent part of the Lagrangian obtained in Eq.\eqref{equivalence action to describe the system}. It is notable that it is indeed of form of the general action in Eq.\eqref{general 2n band Lagrangian we calculate}. Explicitly, we have the following Lagrangian in terms of the constituent fields $\psi=(\phi,\chi)^T$:  
\be{separation of high and low energy part in the Lagrangian with both vector and axial potential}
{\cal L}_{(3+1)d}
=\phi^\dagger(i\bar D_0)\phi+\chi^\dagger(i\bar D_0+2m)\chi
+\chi^\dagger(i\sigma^i{\bar D}_i)\phi
+\phi^\dagger(i\sigma^i{\bar D}_i)\chi
\ee
By choosing $\gamma^\mu$ and $\gamma^5$ as
\be{choosing the correct gamma matrix form in (3+1)d Dirac model}
\gamma^0=\begin{pmatrix} \mathbb{I}_2&0 \\ 0&-\mathbb{I}_2\end{pmatrix},   
\gamma^i=\begin{pmatrix} 0& \sigma^i \\ -\sigma^i&0 \end{pmatrix},
\gamma^5=\begin{pmatrix} 0&\mathbb{I}_2 \\ \mathbb{I}_2&0 \end{pmatrix}
\ee
to express ${\cal L}_{(3+1)d}$ in the form of Eq.\eqref{separation of high and low energy part in the Lagrangian with both vector and axial potential}, the modified covariant derivative with axial potential ${A^5_\mu}$ is given by
\be{definition of bar D_0 and bar D_i}
\bar D_0&=&\partial_0+ieA_0+i\eta\sigma^i {A^5_i}\\
\bar D_i&=&\partial_i+ieA_i-\frac{i}{3}\eta\sigma_i {A^5_0}
\ee
Here we denote vector potential charge as $e$ and axial potential charge as $\eta$. From Eq.\eqref{L2n} we choose ${\cal D}_0=\bar D_0\nonumber$ and ${\cal D}=i\sigma^i\bar D_i$. Comparing with the general expression in $\hat H_{eff}$ as Eq.\eqref{final expression of general effective Hamiltonian}, we have the effective Hamiltonian in powers of $m^{-1}$: $\hat H_{eff}=\sum_{n=0}^2\hat H_{eff}^{(n)}$
\be{final expression of H^(n)_eff for (3+1)d Dirac Lagrangian with both vector and axial potentials}
\hat H^{(0)}_{eff} &=& eA_0+\eta {A^5_i}\sigma^i\nonumber\\
\hat H^{(1)}_{eff} &=& \frac{1}{2m}[\hat{\bm{\pi}}^2+\eta^2 ({A^5_0})^2]+\frac{1}{2m}[-eB_i+\eta(\hat\pi_i {A^5_0}+{A^5_0}\hat\pi_i)]\sigma^i\nonumber\\
\hat H^{(2)}_{eff} &=& \frac{e}{8m^2}[-(\partial_iE_i)+\epsilon_{ijk}(\hat{\pi}_jE_k-E_j\hat{\pi}_k)\sigma^i]-\left(\frac{\eta}{4m^2}\right)(B_{5,i}\hat{\pi}_i)\nonumber\\
&-&\left(\frac{\eta}{8m^2}\right)[(\partial_t\partial_j {A^5_0})+(\hat\pi_i^2{A^5_j}+2\hat\pi_i{A^5_j}\hat\pi_i+{A^5_j}\hat\pi_i^2)-2(\hat\pi_i{A^5_i}\hat\pi_j+\hat\pi_j {A^5_i}\hat\pi_i)]\sigma^j\nonumber\\
&+&\frac{\eta^2}{4m^2}[\epsilon_{ijk}(\partial_i{A^5_0}){A^5_j}\sigma^k]+\frac{e\eta}{2m^2}(B_i{A^5_i})
\ee
where we define $\hat{\pi}_i\equiv\hat p_i+eA_i$. In the limit $A_\mu^5\rightarrow 0$, the generalized Hamiltonian reduces to the conventional Breit Hamiltonian
\be{recall the original Breit Hamiltonian}
\hat H_{\mathrm{Breit}}=eA_0&+&\frac{1}{2m}(\hat{\bm{\pi}}^2-e\mathbf{B}\cdot\bm{\sigma})+\frac{e}{8m^2}[-\bm{\nabla}\cdot\mathbf{E}+(\bm{\hat\pi}\times\mathbf{E}-\mathbf{E}\times\bm{\hat\pi})\cdot\bm{\sigma}]\label{BreitH}
\ee
At the same time, the presence of axial gauge fields leads to additional corrections arising from the combined effects of the mass gap, the vector potential, and the axial potential. These terms are absent in the standard electromagnetic Breit Hamiltonian. A similiar form of Hamiltonian was obtained in Ref.~\onlinecite{SHAPIRO2002113} using the Foldy--Wouthuysen transformation in the context of spacetime torsion. In condensed matter systems, however, dynamical axial potentials can be generated more naturally, for example through strain \cite{Cortijo2015,Pikulin2016,Gorbar20172}, acoustic driving \cite{Rinkel2017,Peri2019}, or photonic means \cite{Hongwei2019}.

\subsection{The Breit Hamiltonian for $(2+1)d$ gapped Dirac systems}\label{Discussion}

The general expression in Eq.~\eqref{final expression of general effective Hamiltonian} can be applied straightforwardly to a variety of gapped fermionic systems. As a useful comparison with the $(3+1)d$  case, we consider a gapped $(2+1)d$ Dirac system. In two spatial dimensions there is no independent $\gamma^5$ matrix. Instead, as in graphene-like systems, the role of chirality is replaced by a valley degree of freedom $\tau\in\left\{+,-\right\}$. The corresponding massive Dirac Lagrangian can be written as
\be{2+1 massive Dirac Lagrangian at tau}
{\cal L}_{(2+1)d}=\sum_{\tau}{\cal L}_{\tau}=\sum_{\tau}\bar\psi_{\tau}[i\gamma^\mu_\tau(\partial_\mu+iA_\mu+i\tau {A^5_\mu})-m]\psi_\tau
\ee
Here we choose $\gamma^\mu_\tau=(\sigma^3,i\tau\sigma^1,i\sigma^2)$ satisfying $[\gamma^\mu_\tau,\gamma^\nu_\tau]_{-}=2\mathrm{diag}(+,-,-)$ and $\psi_\tau=e^{-im t}(\phi_\tau,\chi_\tau)^\mathrm{T}$. Setting $D_\mu^{(\tau)}=\partial_\mu+i(A_\mu+\tau {A^5_\mu})$, we expand Eq.\eqref{final expression of general effective Hamiltonian} to obtain \cite{YwChang2021}
\be{final expression of H_eff for (2+1)-dimension Dirac Lagrangian}
\hat H_{eff,\tau}^{(0)}&=&A_0+\tau {A^5_0}\nonumber\\
\hat H_{eff,\tau}^{(1)}&=&\frac{1}{2m}[(\hat{\bm{\pi}}_\tau)^2+\tau F_{12}^{(\tau)}]\nonumber\\
\hat H_{eff,\tau}^{(2)}&=&\frac{1}{8m^2}\{(\partial_1F_{01}^{(\tau)})+(\partial_2F_{02}^{(\tau)})+\tau([\hat\pi_{2,\tau},F_{01}^{(\tau)}]_{+}-[\hat\pi_{1,\tau},F_{02}^{(\tau)}]_{+})\}
\ee
where $\hat{\bm{\pi}}_\tau=-i\bm{\nabla}+\mathbf{A}+\tau\mathbf{A}^5$ and $F_{\mu\nu}^{(\tau)}=\partial_\mu(A_\nu+\tau {A^5_\nu})-(\mu\leftrightarrow\nu)$.

The $(2+1)d$ result highlights an important distinction from the $(3+1)d$ case. In $(2+1)d$, the two valleys remain independent: the mass term does not mix the $\tau=+$ and $\tau=-$ sectors. Consequently, the Lagrangian can be decomposed into two separate valley contributions, ${\cal L}_+$ and ${\cal L}_-$, and the vector and axial potentials enter only through the valley-dependent combination $A_\mu+\tau A_\mu^5$. This makes the Breit-type Hamiltonian considerably simpler. By contrast, in $(3+1)d$ the mass term mixes left- and right-handed Weyl fermions. As a result, the vector and axial potentials cannot be combined into a simple mass-independent gauge field, and the large-mass expansion produces nontrivial vector--axial couplings in the effective Hamiltonian.

\section{Discussion and Summary}\label{discussionandsummary}

The path-integral technique developed in this work provides a general and systematic route to Breit-type Hamiltonians for gapped fermionic systems. Its usefulness lies in the fact that it does not rely on a model-specific sequence of Foldy--Wouthuysen transformations. Instead, once the low- and high-energy sectors are identified, the high-energy component can be integrated out directly, and the resulting effective theory can be organized as an expansion in inverse powers of the gap. The same procedure applies to general {\color{black}Dirac-type} $2N$-band systems, independent of the detailed microscopic origin of the gap or the explicit dimensionality of the model.

An important feature of this method is that it produces not merely an effective action, but a canonical low-energy Hamiltonian. After the high-energy modes are integrated out, the low-energy field generally acquires a nontrivial symplectic structure. The field rescaling dictated by this structure is therefore essential for obtaining a conventional Schr\"odinger description. This step makes the method especially useful in condensed matter applications, where Hamiltonian formulations are often the most convenient starting point for studying spectra, wave-packet dynamics, and transport phenomena.

The generalized Breit-type Hamiltonian derived here contains several terms whose physical implications are not immediately transparent from their operator form in Eq.~\eqref{final expression of H^(n)_eff for (3+1)d Dirac Lagrangian with both vector and axial potentials}. In our companion paper~\cite{shortpaper}, we use this Hamiltonian as the starting point for a semiclassical wave-packet Boltzmann formalism and apply it to transport phenomena in Weyl excitonic insulators. This illustrates how the general formalism developed here can be translated into experimentally relevant predictions.
To highlight one consequence, consider the ${\eta[\hat{\pi}_i,A^5_0]_{+}\sigma^i/2m}$ term in $\hat{H}_{eff}^{(1)}$ of Eq.~\eqref{final expression of H^(n)_eff for (3+1)d Dirac Lagrangian with both vector and axial potentials}. This term originates from an imbalance of chemical potentials between the original Weyl nodes \textcolor{black}{and the energy separation bewteen two Weyl nodes}, represented by a nonzero axial scalar potential $A^5_0$. In the gapped phase, it acts as an effective spin-orbit coupling for the low-energy quasiparticles and generates a nontrivial Berry connection. As a result, a dynamical axial potential can induce a Hall-type transport current in Weyl excitonic insulators.

In summary, we have developed a path-integral framework for deriving Breit-type effective Hamiltonians for general $2N$-band gapped {\color{black}Dirac-type} systems. The method reproduces the conventional Breit Hamiltonian for the ordinary Dirac equation and naturally extends to systems with axial gauge fields, where it generates additional vector--axial couplings. The comparison between $(2+1)d$ and $(3+1)d$ systems further shows that mass-induced mixing of left- and right-handed fermions in three dimensions leads to genuinely new low-energy terms. These results establish the path-integral construction as a general and versatile tool for uncovering low-energy physics in gapped Dirac and Weyl materials.

\section{Acknowledgments}
This work was supported by National Natural Science Foundation of China (NSFC) under Grant No. 12374332, Quantum Science and Technology-National Science and Technology Major Project No. 2021ZD0301900, Cultivation Project of Shanghai Research Center for Quantum Sciences Grant No.LZPY2024, and Shanghai Science and Technology Innovation Action Plan Grant No. 24LZ1400800.

\bibliography{ref_longpaper}
\let\clearpage\relax
\end{document}